\documentclass{IEEEtran}
\usepackage{cite}
\usepackage{placeins}
\usepackage[pdftex]{graphicx}
\graphicspath{{./png}}  
\usepackage[cmex10]{amsmath}
\usepackage{amssymb}

\begin{document}
\title{Increasing the Capability of Neural Networks for Surface Reconstruction from \\ Noisy Point Clouds}
\author{Adam~R~White, Li Bai \\ School of Computer Science \\ University of Nottingham \\ Jubilee Campus, Nottingham NG8 1BB, UK}

\maketitle

\begin{abstract}
3D modeling from point cloud data is not exclusively in the vein of aesthetics but industrial products and medical scanners. With the emergence of technologies like VR, and the reducing cost of personal laser scanners, the need for algorithms that give realistic representation of scanned data is ubiquitous. This paper builds upon the current methods to increase their capability and automation for 3D surface construction from noisy and potentially sparse point clouds. It presents an analysis of an artificial neural network surface regression and mapping method, describing caveats, improvements and justification for the different approach. 
\end{abstract}

\section{Introduction}
Accurate surface reconstruction from noisy point cloud data is still an unsolved challenge. Raw point cloud data is unstructured and may be noisy or sparse. The challenge can be tackled with a neural network (NN) approach \cite{Yumer and Kara (2011)}, to learn a mapping from a 2D parameterisation of 3D point cloud data, resulting in a surface that is less sensitive to noise. This approach differs to standard NN applications as both the hyperparameters and intrinsic parameters change during training in order to find the optimum model \cite{Yumer and Kara (2011)}. 

This paper considers the 2D parameterisation method, directly comparing the current dimension reduction used and other familiar methods. A least squares Spline fitting is applied to define and interpolate the boundaries of the surface mesh. The interpolants are carefully chosen by a method presented to enable a fit that is more faithful to the boundary of objects. A working implementation is demonstrated, producing good quantitative and qualitative results on a range of noisy datasets. 

\section{Related Work}
An approach was suggested by Peng et al \cite{Peng et al (2001)} using an image processing inspiration for surface de-noising. It removes noise by applying a Weiner filter which approximates the components of the surface as a statistical distribution. There are two problems with this algorithm in our context. First, it needs to be decided \textit{when} this algorithm should be applied, unnecessary smoothing might remove features that describe the underlying geometry, although there is some attempt to apply a surface based anisotropic diffusion to preserve edges. In addition, the formula used requires the user to both know and supply the noise denoted by the variance $\sigma^2$ \cite{Peng et al (2001)}. It may not be possible to determine the noise of the data as it is an unstructured point cloud.

Mederos et al attempts to find local approximations of implicit surfaces on an object which is then combined into a global description \cite{Mederos et al (2007)}. This algorithm uses an octree to recursively divide the input space according to an error measured at a node. With Machine Learning in context an interesting method was proposed where the knot vectors and control points of B-spline curves and surfaces are learned \cite{Van To and Kositviwat (2005)}. 

Yumer and Kara suggest a NN regression method of surface fitting and hole stitching. The flexibility 
achieved by an adaptive neural network topology differs from previous attempts as the ideal topology of the 
network obtained (the hyperaparmeters) are not fixed \cite{Yumer and Kara (2011)}, meaning the network can be tailored to each point cloud automatically. This method is good for removing noise as the underlying geometry of the point
data and not random noise is represented in the final surface.

In a slightly different problem, where a NN is used to reconstruct the shape of a 3D object from its shading in a 2D \cite{Khan et al (2009)}. Khan et al show from experiment that quantitative improvement does not necessarily lead to quantitative improvement. This is something to consider when using a 'black box' function like a neural network, especially where there could be some information loss. In this regard we must ensure that the final model is representative of the ground truth and not only rely on an error measure. It is suggested that more research must be done for 3D surface quality metrics \cite{Khan et al (2009)}. Visual quality will be assessed in the method presented here alongside quantitative results in the absence of quality metrics.

Many papers in the field use a least squares based optimisation to curve or basis fitting \cite{WANG et al (2006)}
\cite{Dierckx (1993), Caitlin et al (2015)} to cite a few. Wang et al (2006) admits that this automated fitting is still a problem in graphics \cite{WANG et al (2006)}. They introduce a direct SD squared distance error function which is iteratively updated by using quasi-Netwon gradient descent. The spline in this case has an initial shape which is fit to the point cloud by taking the 1st order derivative of the objective function. The method used in their paper assumes the knots are fixed in place \cite{WANG et al (2006)}. The method chosen for our research allows variable knots
as this allows for a better fit and flexibility of model \cite{Dierckx (1993)}. Despite this constraint the paper shows excellent empirical results and would have been tested had an implementation been readily available for the programming language used for this paper.

A recent development is the use of Spherical Harmonics for Modeling where a surface can be described by three spherical functions based on a bijective between the Cartesian (x,y,z) and $(\theta, \gamma)$ the spherical domain \cite{Shen et al (2008)}. The method by Caitlin et al \cite{Caitlin et al (2015)} transform mesh vertices into spherical coordinates and use a form of Tikhonov regularisation to create a smooth mesh of the surface. While our solution does not take the same form, we consider a smoothing criteria similar to Caitlin et al (2015).

\section{Materials and Methods}
We aim to create an accurate surface from noisy point cloud data by analysing and improving a NN approach by \cite{Yumer and Kara (2011)}. The steps required to achieve goal of this paper are:
\begin{enumerate}
	\item Use the Isomap algorithm to reduce the dimensions of the point cloud $\mathbb{R}^3 \rightarrow \mathbb{R}^2$
	\item Train NN to map $\mathbb{R}^2 \rightarrow \mathbb{R}^3$. This mapping learned by sampling the initial
		point cloud and using the points as training and test data.
	\item Use a multi-depth path method to choose the outer-most points
		of the 2D manifold to be interpolated
	\item Least Squares fit a cubic B-spline to the interpolants chosen with a justifiable choice
		of regularisation 
	\item Re-sample points inside of the boundary dictated by B-spline
	\item Find the triangular tessellation of the point cloud by Delaunay triangulation 
	\item Feed NN the points in  $\mathbb{R}^2$ to produce the target vertices in $\mathbb{R}^3$
	\item Mesh output according to triangular topology produced in $\mathbb{R}^2$
\end{enumerate}

\subsection{Feature Selection and Dimensional Reduction}
The purpose of dimension reduction in this case is to allow a NN to learn a mapping between the 2D coordinates and the 3D coordinates \cite{Yumer and Kara (2011)}. Also, it simplifies the problem of surface generation: The boundary of the point set can be easily defined in 2D and the topology of the mesh established. Later the vertices of the mesh are fed to a trained NN. 

The first step of the proposed algorithm is to embed the points in $\mathbb{R}^3$ to $\mathbb{R}^2$. Our feature space will only ever be $\mathbb{R}^3$, as is intrinsic to generating 3D surfaces in Euclidean space. To facilitate the 2D embedding the use of Isomap algorithm is suggested, originally proposed by Tenenbaum et al in 2000 \cite{Tenenbaum et al (2000)}. The holistic reason for using Isomap over other dimension reduction algorithms is because Isomap intends to preserve global geometry \cite{Tenenbaum et al (2000)}. Given the goal is to extract the underlying geometry and not the noise of the point cloud to produce a smooth surface, the 2D embedding must be representative.

The hallmark of Isomap is that points are reconstructed according to their pairwise geodesic distance. A graph for each neighborhood is used to represent the distance path where each edge is weighted, usually by euclidean distance. A neighborhood is defined by either by K points or a radius denoted by $\sigma$.
\begin{center}
	\begin{enumerate} % DIRECT QUOTES?
	\item Determination of neighborhood for each pair of points j,i is given by $d_x(i,j)$ and store
	relations as a weighted graph G. \\
	\item Compute the shortest path distance $d_G(i,j)$ using an algorithm such as Floyd-Warshall. \\
	Once the graph distances are obtained as matrix $D_G = {d_G(i,i)}$ the Multi-dimensional Scaling
	algorithm is applied. \\ 
	\item Finally the coordinate vectors in the resulting space Y are reconstructed to reduce the cost function: \\
	\end{enumerate}
	\begin{equation} E = ||\tau(D_G)-\tau(D_Y)||_{L^2} \end{equation} from \cite{Tenenbaum et al (2000)}
	$ D_Y $ represents the matrix of Euclidean distances. \\
	$ \tau $ converts the distances to inner products \\
\end{center}

\subsection{Comparison}

\subsubsection{Qualitative Comparison}
To make our investigation more critical we compare Isomap and Locally Linear Embedding (LLE) \cite{Yumer and Kara (2011)}. Silvia and Tenenbaum, the original authors behind Isomap, published a comparison between Isomap and LLE two years earlier. In the defence of LLE it was suggested that it would be useful on a broader range of data when the local topology is close to Euclidean \cite{Silvia and Tenenbaum (2002)}. LLE is attractive in this regard as many surfaces to be produced will have local geometry that is close to Euclidean. We also desire a representative 2D embedding that is conducive to an accurate output once input to the learned NN. When noise is present it is more important that the general structure of the point cloud is learned and not the noise. Global methods, like Isomap, tend to give a more 'faithful' representation with respect to its global geometry \cite{Silvia and Tenenbaum (2002)}. With two algorithms, and two desired properties, we attempt to evaluate (and decide on the best) applicability to the niche problem in this section.

\vspace{0.1cm}
\FloatBarrier
\begin{figure}[h]
\centering
\includegraphics[width=0.5\linewidth]{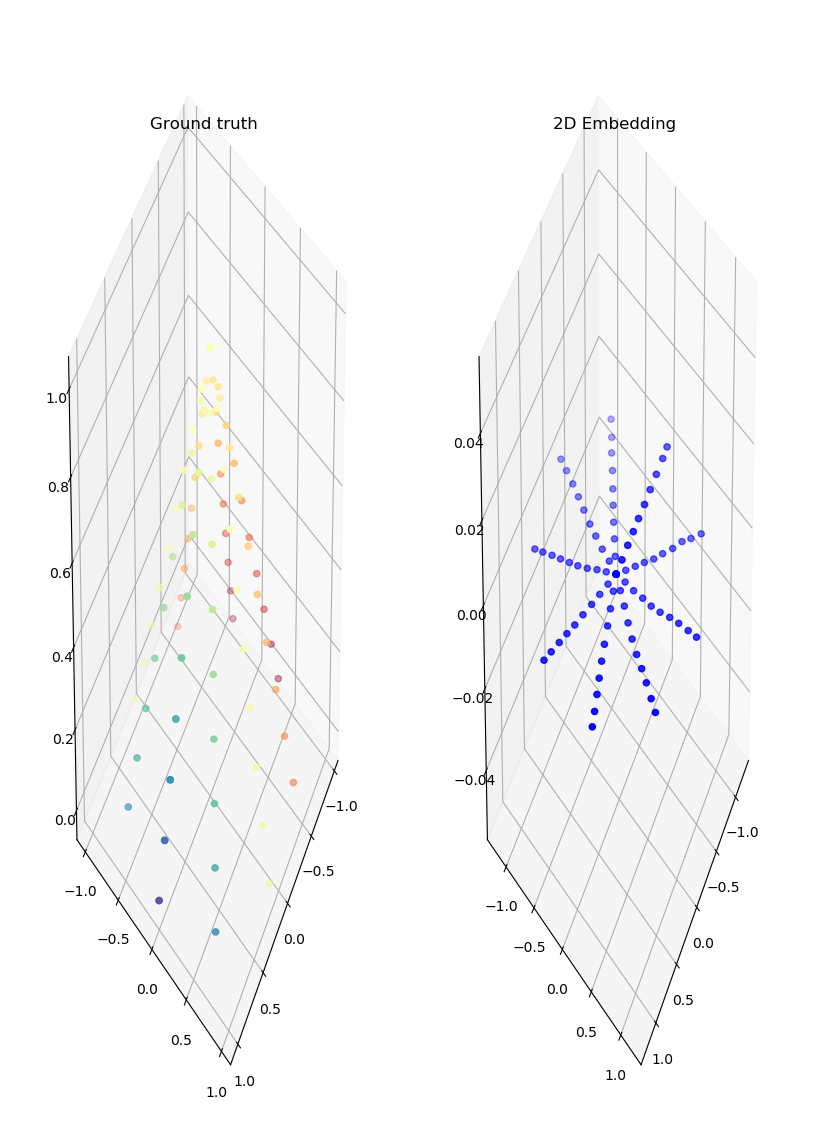}
\caption{Isomap with 12 nearest neighbors}
\end{figure}
\begin{figure}[!h]
\centering
\includegraphics[width=0.5\linewidth]{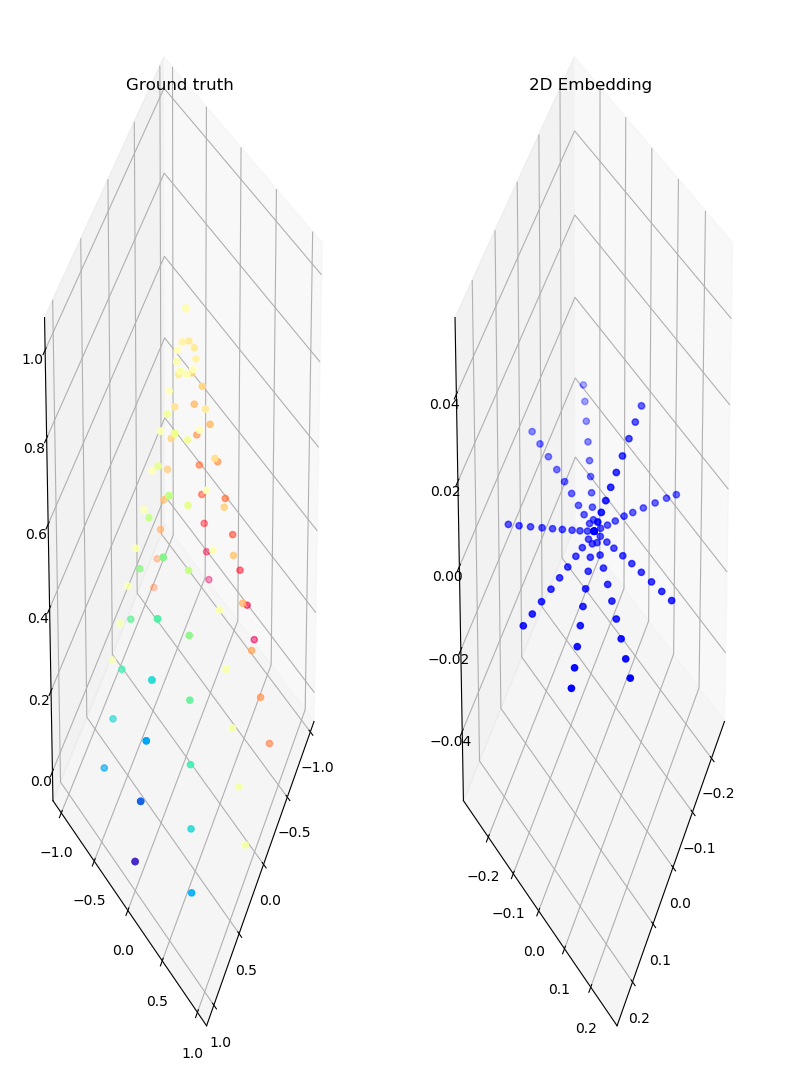}
\caption{LLE modified weight method \cite{Zhang and Wang (2006)} 12 nearest neighbors}
\end{figure}
\FloatBarrier
Before a surface was constructed we compared the output of the trained NN against the original data (the Standford Bunny point cloud with 1600 points) to ascertain which two dimensional input gives us results closest to the original. All methods gave expected results on a more complex and dense point cloud. The Hessian eigenmap method by Donoho and Grime \cite{Donoho and Grimes (2003)} performs poorly on capturing the relative scale of the ears. The modified weight \cite{Zhang and Wang (2006)} LLE method seems to give the most intuitive results.

\subsubsection{Quantitative Comparison}
% Compare ISOMAP!
\begin{figure}
\centering
 \begin{tabular}{|l |l |l |l| l| l|}
	 \hline 
	 IO layers & Method & K & MSE r\\ \hline
	 Sigmoid & Isomap & 12 &  0.001503\\ \hline
	 Sigmoid & Mod LLE & 12 & 0.001483\\ \hline
	 Linear & Isomap & 12 & 0.0003200 \\ \hline
	 Linear & Mod LLE & 12 & 0.0004637 \\ \hline
 \end{tabular}
 
	\caption{All other constraints on the neural network: epochs, max layers, max neurons were fixed. }
\end{figure}
\vspace{1cm}
We conducted further systematic tests to distinguish the proper use of either a global geodesic method like Isomap or the Modified LLE. In order to ensure results weren't reflective of a particular dataset or neural network topology, all combinations of activation functions from a pool of well known functions were chosen and tested.

\begin{figure}[h]
\centering
\begin{tabular}{|l |l |l |l| l| l|}
	 \hline 
	 Data & Method  & MSE & Points\\ \hline
	 S-curve & Isomap & 0.1148 & 400\\ \hline
	 S-curve & Mod LLE & 0.2665 & 400 \\ \hline
	 Torus & Mod LLE & 0.02119 & 100 \\ \hline % is it the number of points that makes a difference
	 Torus & Isomap & 0.03138 & 100\\ \hline
	 Sparse Cone & Mod LLE & 0.03218 & 36 \\ \hline
	 Sparse Cone & Isomap &  0.04621 & 36 \\ \hline
	 Denser Cone & Isomap &  0.02240 & 144 \\ \hline
	 Denser Cone & Mod LLE & 0.09032 & 144 \\ \hline
\end{tabular} 
\caption{Max Layers : 3, Max neurons : 6 Epochs : 20 Early Stop : 3 MSE (Mean Squared Error)}
\end{figure}
The size constraints of the network were kept relatively small to account for a lengthy training+test run time.

Both methods are dependent on the activation functions and both methods give similar qualitative results as error. However, it became apparent that a second unintended independent variable, being the cardinality of the points in the data set, affects the final NN output error for different methods of 2D embedding. The results show that the Modified LLE method has the edge for very sparse data whereas Isomap gave better results on denser (relatively speaking) datasets. Modified LLE often failed to run at all on dense datasets and we have discounted the traditional method of LLE especially given its non deterministic output. Therefore, from the experiments conducted, the use of Isomap for this problem is suggested, unless its known in advance that very sparse data will be used.

Isomap, however, is not perfect. One problem of Isomap, that the next stage of our method tries to mitigate, is that
it often highlights outliers. Outliers from noise that occur outside the manifold may be included in the transformation
of the points from 3D to 2D. Both methods suffer the problem of incorrectly choosing 'K' (points in the neighborhood) and lead to poor results. This method employs no heuristic to choose an optimum K. Extreme error values may indicate 
that the value of K should change however a more concrete system of heuristics must be included to reduce user interaction. 

\section{Training a neural network}
A neural network is used to learn the mapping between our embedded 2D points and the ground truth 3D point cloud. We use a NN in this context as it is hoped this interpolation property captures the general structure of the point cloud and not the noise. Noisy data is approximated by a linear regressor function, resulting in a smooth approximation of the underlying distribution, thus avoiding the scattered non-uniform raw data. Given any point cloud, the NN can fit a function to the noisy data that can represent any general function. This property is most desirable as it implies the NN method can be applied to a huge range of different data. The final form of the whole network as shown for use in \cite{Yumer and Kara (2011)}.

\begin{equation}
{\overrightarrow{D}}_k = f(\sum^{n}_{j=1}w_{kj}f(\sum^{2}_{i=1}w_{ji}{\overrightarrow{P}}_i+w_j0) + w_k0) 
\end{equation}
Where f is to be decided. 

The method of training builds on that of Yumer and Kara's, and this is the only part of the work that has not not been changed in an important way.

\begin{enumerate}
	\item Segment the input data into random samples of 85\% training, 10\% test, 5\% validation, 
	\item Initialize a network with a single hidden layer and a single neuron
	\item Train the network until the validation set performance converges, with back-propagation
		and early stopping to prevent overfitting
	\item Record the weighted training-test set performance for the current network configuration 
	\item Increase the number of hidden neurons by 1. Iterate steps 3-5 until the weighted performance
		converges or the number of neurons reaches a maximum 
	\item Record the number of neurons and the test performance for the current layer
	\item Iterate steps 3-7 until the number of layers reaches a maximum 
	\item Return the network configuration with the best weighted performance
\end{enumerate} \cite{Yumer and Kara (2011)}

\section{Surface generation}

\subsection{Defining the manifold}
Once the point set is embedded in two dimensions the next task is to sample the edge points of the now 2D point cloud and fit an curve to define the manifold. Once the 2D vertices generated by the procedure are fed to the trained NN the resulting points in 3D represent less noisy version of the intended surface. The output points of the NN become vertices of a triangular mesh. 

Prior research makes the assumption that the boundary point set of the 2D embedding is a reasonable outline of the expected shape, but with very noisy data outliners outside the expected boundary would still be considered as valid points, causing the manifold to appear perturbed. We use the idea of re-sampling the inner points with a regular grid, based on Yumer and Kara's work \cite{Yumer and Kara (2011)}. The method chosen here is outlined below. It makes no assumptions about the quality of the boundary points and can handle outlier points not representative of the manifold.

\begin{enumerate}
\item Sample a proportion of the outer most points in the cloud using the multi-line sampling method described in 'Choosing the Interpolants' 
\vspace{0cm}
\item Use sampled points to fit a cubic B-spline curve using Least Squares fit with regularisation \cite{Dierckx (1993)}
\item Superimpose regular grid on point set and uniformly re-sample points inside of B-spline loop
\end{enumerate} 

In order to represent the outline of the 2D point cloud we use a cubic B-spline curve to better fit the local boundary of the dataset \cite{Vince (2014)}. The B-spline is defined as follows:

\begin{equation} 
\textbf{S}_i(t) = \sum_{r=0}^{3} \textbf{P}_{i+r}B^k_r(t) 
\hspace{1cm} \text{ for } 0 \leq t \leq 1 \end{equation}
\textit{where r denotes blending ratio and k the degree of the Bernstein basis}

\begin{equation}
B_i^1 = \left(\frac{t-t_i}{t_{i+1}-t_i}\right)+\left(\frac{t_{i+2}-t}{t_{i+2}-t_{i+1}}\right) 
\end{equation}
\textit{when $t_i \leq t \leq t_{i+1}$ the Berstein basis and associated control point blend in. \\ when $t_i \leq t \leq t_{i+2}$ the control point and Bernstein basis blend out}.

\vspace{0.5cm}
The task of deciding which $p_i$ control points, the sequence of values for $t_i$ (knot vector) and additional weights
that interpolate the points best will be discussed in surface fitting.

\subsection{Choosing the interpolants}
For the purposes of fitting the B-spline we are reducing the problem to a Least Squares interpolation, and the interpolation is only as good as the interpolants are representative. In similar work it was suggested that the outermost path connected by 4 corners by polyline \cite{Yumer and Kara (2011)}. This works just fine for relatively noise free data but it became apparent that the method can be improved for noisy data sets and made more precise for non-noisy data. We can not use a 1-point deep outer loop as the probability of this being true to the noise free representation of the surface outline is very low. With the addition of the noise, the outer-loop will very likely be perturbed. However, unless the points do not even remotely resemble the expected geometry, we do know that somewhere between the outer most points and a few points in the centroid direction lies the outline of the 'perfect' noise free shape. 

\begin{itemize}
	\item For each depth 
	\item Pick 8 (or more if desired) corners according the furthest distance in circular sector from the centroid
	\item For each corner
	\begin{itemize}
		\item Segment point space into rectangles containing all 	points between one corner and its adjacent corner
		\item Consider points between as weighted graph (using KDTree
			in this case)
		\item Set weights to be $ w = c_1.d_s + c_2.d_c $ where $d_s, d_c$ are the straight line
			distance to adjacent point and centroid, respectively. c defining the 'importance'
		\item Find the shortest path to adjacent point using w as criteria for next point
			selection
	\end{itemize}
	\item Strip path containing selected points away from cloud so as not to be calculated for next depth
	\item Combine path returned for all corners 
\end{itemize}
\vspace{0.5cm}
This method differs from previous attempts as more 'anchor' points are selected according to the spoke sampling criteria mentioned earlier. The incorporation of more anchor points gave better local precision for finding a path as the distance, and thus points considered between each anchor, ensures the algorithm has less room for error: by a short circuiting path, for example. Further, our method allows adjustable depth of points to be sampled so more interpolants can be considered when fitting a boundary.
%\FloatBarrier

\begin{figure}[h]
 \centering
	\includegraphics[width=.6\linewidth]{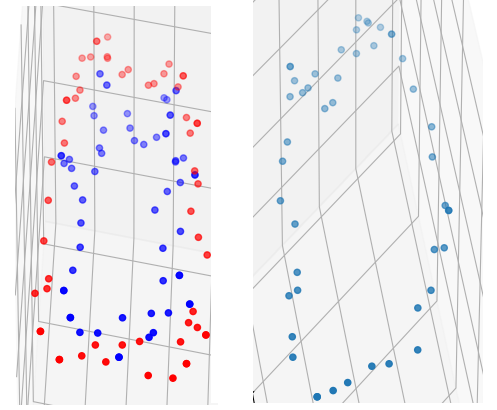}
	\caption{Our Multi-depth bath-based sampling with noise (left).
	Polyline samples with PCA corners similar to \cite{Yumer and Kara (2011)} (right)
	You can see how easily the second method is susceptible to noise. Even if re-sampled.}
\end{figure}
\FloatBarrier

\subsection{Surface fitting}
In order to fit the spline to the selected interpolants we use Dierckx's algorithm for least squares fitting a B-spline with variable knot vectors. Where the knot vector is a vector of initial values for the 'blending ratio' and define the amount of 'blending' for each control point on the curve (equation 8). The general form of the least squares for the fitting a B-spline is arranged by Dierckx as:

\begin{equation} \delta = \sum_{r=1}^{m}\left(w_ry_r-\sum_{i=-k}^{g}p_iw_rB_i^{k+1}(x_r)\right)^2 \end{equation} from \cite{Dierckx (1993)}
\begin{itemize}
    \item data points: $(x_r, y_r)$
    \item set of weights $w_r$
    \item the control points $p_i$
    \item the number an position of the knots \textbf{t}
    \item substitute the blending ratio and Bernstein basis '$B_i^k$' from (4).
\end{itemize}
We have only shown a system where the knot vectors are fixed. Here, we will keep the description for picking the appropriate knots brief and suggest readers seek \cite{Dierckx (1993)} for a more vigorous explanation. Dierckx avoids the problem of coinciding knots, and the existence of knots very near to the basis boundary [a,b], by separating (3) the least squares spline objective function and penalising the overall error using the following heuristic.

\begin{equation}
	\epsilon(\textbf{t}) = \sigma(\textbf{t})+pP(\textbf{t}) 
\end{equation} from \cite{Dierckx (1993)}
\textit{where p is not to be confused with a control point and is set according to some heuristic}
\begin{equation}
	P(\textbf{t}) = \sum_{i=0}^{g}(t_{i+1} - t_i)^{-1}
\end{equation} from \cite{Dierckx (1993)}
It's plain to see that the penalty is inversely proportional to the 'closeness' of two adjacent knots. Due to local stopping points on the boundary \cite{Dierckx (1993)} these constraints help avoid poor gradient based minimisation. 

With the residual error $\sigma$ from equation 5. Dierckx suggests the objective function can be subject to the 
constraint that $ \sigma \leq s $  where s is a user selected constant \cite{Dierckx (1993)}. This allows the error function some flexibility so that the spline is not forced to traverse every point exactly. For the agenda of this 
paper a smooth fit is highly desirable. To this end, we attempt to build upon the smoothing property introduced 
by Dierckx. Picking the value of 's' is the most challenging task. While there exists some heuristics to setting 's' before fitting the spline, there was improvement on setting an arbitrary regularisation term by using the variance of y values in the target point set.

\begin{equation}
	\delta \leq \lambda \frac{\sum_i^n(y_i-\hat{y})^2}{|y|}
\end{equation}
Given that we want a curve that smoothly traverses the interpolants, the rationale behind using the variance is that the larger the discontinuity of values the greater allowance for fitting error thus the smoother the fit for jagged interpolants. $\lambda $ is set at a default value but can be tweaked should the user desire. The pit fall with this method is that different values of lambda can still be tailored to each dataset thus the selection of the regularisation is not a fully automated process for finding the optimum qualitative results.
\vspace{0.5cm}

One problem with triangulation algorithms is that they are not well suited to concave point clouds and cause edges outside the outer loop of the points to be created. Currently the working implementation for surface generation uses a basic method for removing triangles outside of the boundary defined by the fitted B-spline. The algorithm simply:

\begin{itemize}
	\item Compute the centroid of each triangle in the triangulation
	\item Consider spline as polygon loop which encompasses all correct triangulations 
		by centroid 
	\item Triangles with centroids that exist outside of B-spline polygon are removed
\end{itemize}
Note that this method assumes that an optimal regularisation term has been picked otherwise jagged and overfit boundaries lead to the removal of triangles that would contribute to good definition of the surface.

\section{Results and Observations}

The implementation is written in python and has 3 main dependencies: Pybrain, Sklearn\cite{scikit-learn (2011)} and Matplotlib. 

We show the effect of using a retrained network verses a copy of the best network found in training. If 'Retrained' is 'Yes', this indicates that a new NN was trained on all the data available after the best topology was discovered during training. If 'Retained' was 'No' then an exact copy of the trained network with the best topology during training was used on the whole data set. 'Final Error' reflects the mean squared error of the network output given the whole data set as an input and not just random samples. Its worth noting that the error in this case is compared against the noisy data so a very low error can mean that the points have overfit to the noise. It should also be noted that this wasn't the case for the earlier results in the 'dimensionality reduction comparison' section, where the error was defined against the 'perfect' error free parametric representation of the intended object. \\

Overfitting on the test data needs to be avoided as it is the overall structure of the geometry that must be captured,
and the test set, while sampled randomly, will skew the Final Error if we allow the NN to overfit. The initial preconception that a retrained network, with optimum hyperparameters, would perform better. Having been trained on the whole data set an expected a closer representation of the ground truth, but experimentation shows that a larger number of epochs causes the error of the retrained NN to diverge from the improvement shown by the other non-retrained NNs.

\begin{figure*}[h]
	\centering
	\includegraphics[width=0.8\linewidth]{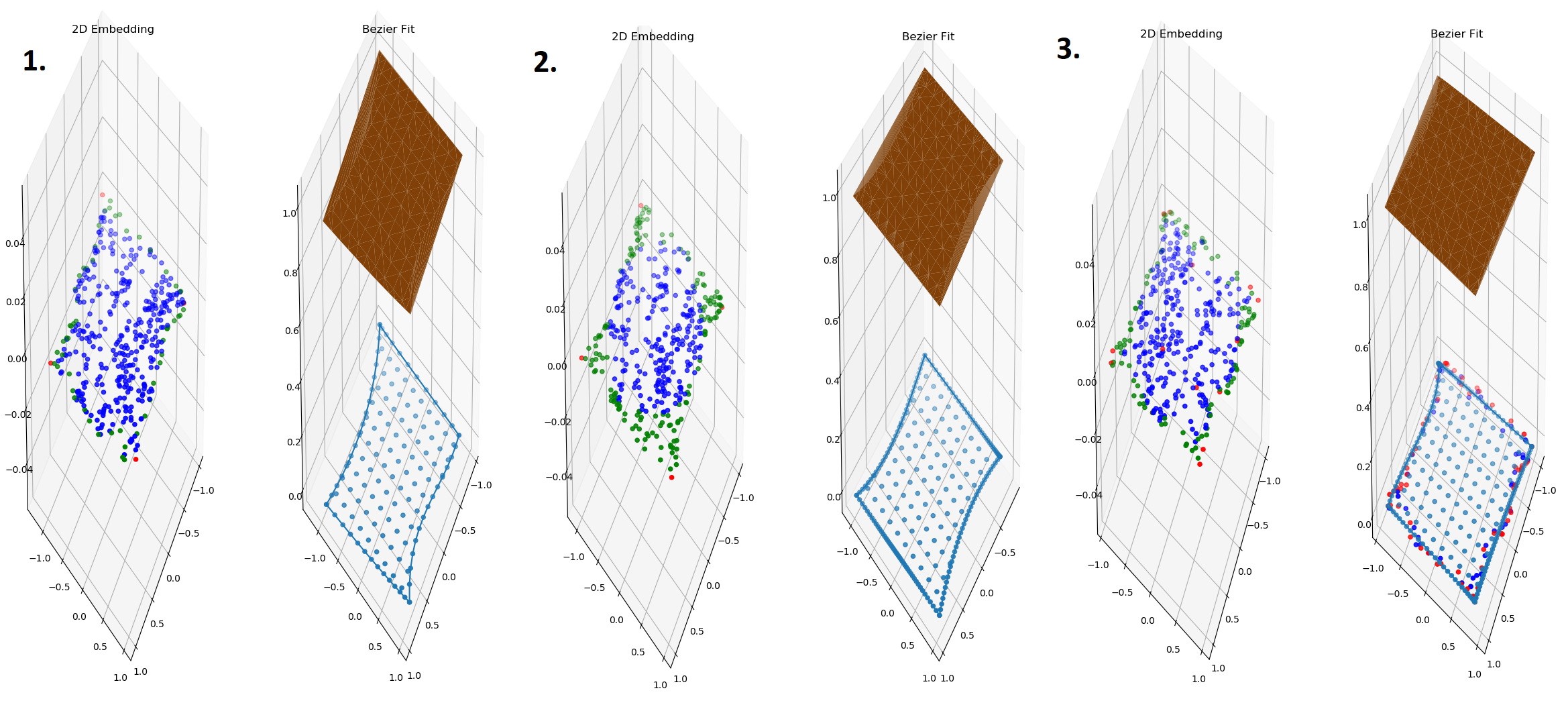}
	\caption{1) PCA polyline sampling, 2) Spoke-like sampling, 3) Sampling method used in this paper}
\end{figure*}
\begin{figure*}[h]
	\includegraphics[width=\linewidth]{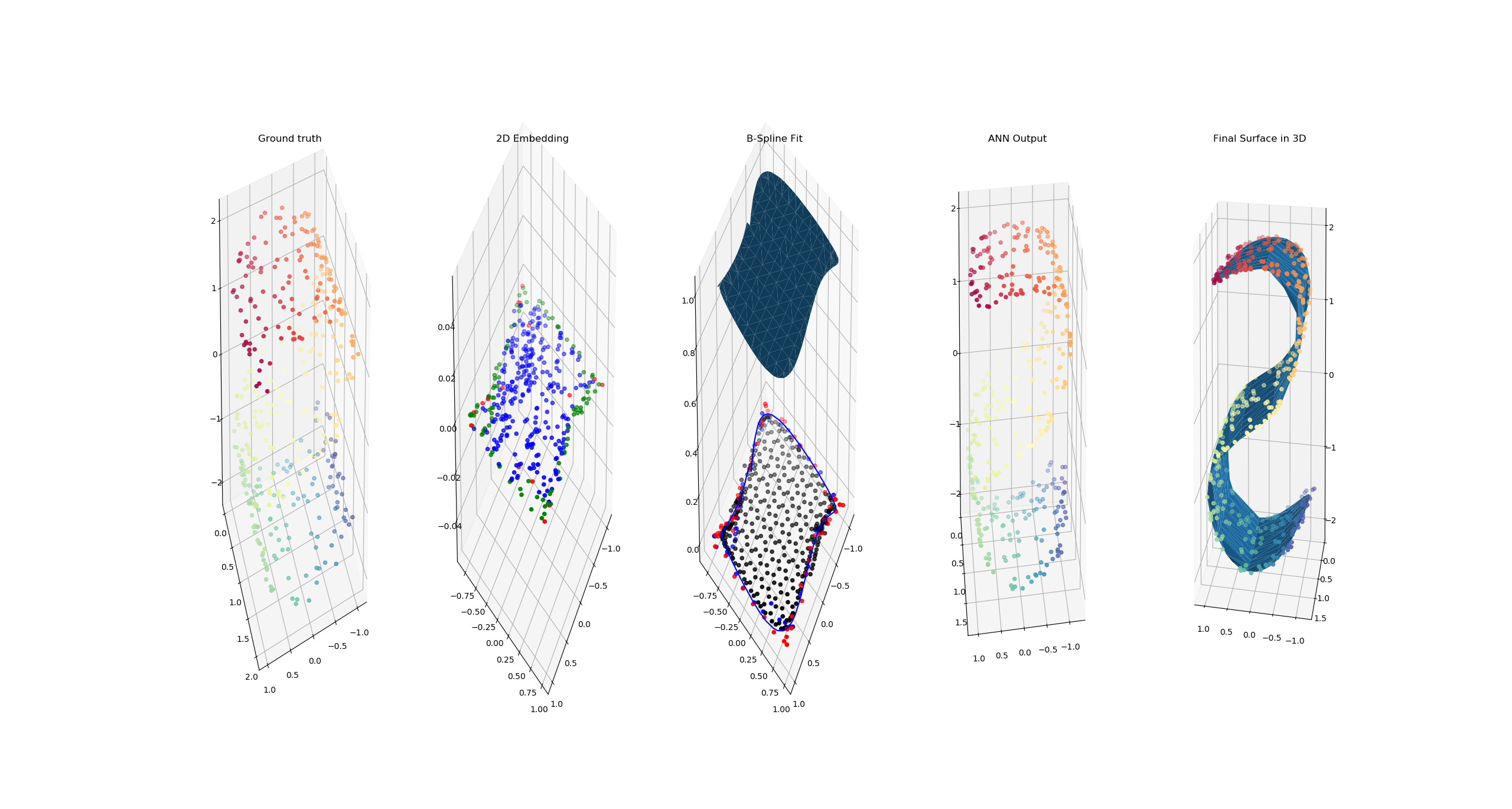}
	\centering
	\caption{Deep Learning potential attempted with 20 neurons max per layer. $\lambda=2.4$}
\end{figure*}
\begin{figure*}
\centering
 \begin{tabular}{|l |l |l |l| l| l|}
	 \hline 
	 Dim Reduction & NN layers & NN neurons & Epochs & Train Error mse & Final Error mse \\ \hline
	 Isomap & 1 & 6 & 100 & 0.006328 & 0.000798 \\ \hline
	 LLE (original) & 1 & 10 & 100 & 1.429 & 0.296 \\ \hline
 \end{tabular}
	\caption{ for $\theta = [0.. \frac{\pi}{2}], \gamma = [0..\frac{\pi}{2}]$ on Torus}
	\hspace{1cm}
\noindent\makebox[\linewidth]{
\begin{tabular}{|l |l |l |l |l | l| l |}
	 \hline 
	 Dim Reduction & NN layers & NN neurons & Epochs & Test Error mse & Final Error mse & Retrained \\ \hline
	 LLE & 1 & 10 & 10 & 0.9102 & 1.071 & Yes\\ \hline 
	 LLE & 2 & (10, 6) & 100 & 0.5376 & 1.212 & Yes \\ \hline 
	 LLE & 3 & (10,10,1) & 10 & 0.7643 & 1.130 & No\\ \hline 
	 LLE & 2 & (10,2) & 100 & 0.3519 & 0.7654 & No \\ \hline 
	 LLE & 2 & (10, 4) & 100 & 0.7191 & 0.6382 & No\\ \hline 
\end{tabular}}	 \caption{for $\theta = [0.. \pi], \gamma = [0..\pi]$ on Torus}
\end{figure*}
\newpage

The quality of the boundary of the point cloud directly affects how perturbed the resulting tessellation will appear on the final output. We will compare other methods of choosing interpolants and show the importance of having more points afforded to interpolation by allowing a variable depth of samples that form the outer loop. In order to keep the independent variables limited to 1, and to show the problem in a simple manner, the following images show the least squares fit of a Bezier curve for simple points sets. This is the same process as \cite{Yumer and Kara (2011)} - what changes is the sampling method:

\section{Conclusion}
Discussed in this paper is the methods for surface generation from noisy point cloud data. Through experimentation we have been able to expose the internals of the algorithms suggested, and give a closer comparative review of this highly specific application of neural networks. The presented method gives good quantitative and qualitative results for a variety of different data sets. With a little more refinement, particularly in the training of the NN, it is hoped that this method can be extended for more complex 3D point clouds.

A software improvement which will speed up the training time of the neural network is the use of a more modern library than Pybrain. On top of this, a better training method should be used to reduce training time. Currently, the training method is simply a gradient descent backpropgation. Not much attention has been payed to parameters like the momentum update, learning rate and other hyperparameters. It would be prudent to refine the method of training as this will be the slowest part of the algorithm. While most of the datasets and neural networks in this, and other papers \cite{Yumer and Kara (2011)}, are constrained to small manageable sizes. The potential for larger, that is deep, with many layers defined in\cite{Goodfellow et al 2016}, should not be overlooked for complex mappings. However, this is only feasible in a reasonable time frame if the selection of the
hyperparamters is more efficient than an exhaustive grid search, otherwise segmentation of the problem will need to be
employed.

Finding the best regularisation value should be also be on the agenda for improvement. The regularisation parameter, while manifold specific, is fixed throughout the fitting of the B-spline. Ideally there need to some variability in the amount of regularisation at the moment of update of least squares. It would be good to the follow a ridge regression trend in further implementations of this algorithm where the regularisation can be more closely dependent on the spline being fit in a similar way as Caitlin et al use the spherical harmonic order to construct the Tikhonov matrix \cite{Caitlin et al (2015)}. As mentioned, regularisation strongly affect the qualitative results and currently a free parameter $lambda$ exists that is not automatically decided.

There needs to be some decision towards when hole fitting is appropriate. This method will blindly re-sample the inside of the point cloud whether or not holes where intended. To be able to realise more complex surfaces with intentional holes segmentation could be used but this immediately increase the complexity of the algorithm. To find a global method which fills in only unintended holes is a very challenging task but would improve this algorithm.

\end{document}